\documentclass[
    aip,apl,
    amsmath,amssymb,
    reprint]{revtex4-2}

\usepackage{graphicx}
\usepackage[utf8]{inputenc}
\usepackage[T1]{fontenc}
\usepackage{mathptmx}
\usepackage{siunitx}
\usepackage{color}
\usepackage{bm}				
\usepackage{mathrsfs}
\usepackage{upgreek}
\usepackage{ORCIDinREVTeX}
\usepackage[hidelinks,
    colorlinks=true,
    linkcolor=blue,             
    citecolor=blue,            
    filecolor=blue,            
    urlcolor=blue]{hyperref}    

\allowdisplaybreaks

\newcommand{\Exp}{\mathrm{e}}
\newcommand{\iomega}{\mathrm{i}\omega}
\newcommand{\im}{\mathrm{i}} 
\newcommand{\nomega}{\widetilde{\omega}}
\newcommand{\inomega}{\mathrm{i}\widetilde{\omega}}
\newcommand{\ep}{\text{EP}}
\newcommand{\identitymat}{\mathbb{I}}
\newcommand{\figref}[1]{\hyperref[{#1}]{\textup{FIG.~\ref*{#1}}}}
\newcommand{\secref}[1]{\hyperref[{#1}]{\textup{Section~\ref*{#1}}}}
\newcommand{\appref}[1]{\hyperref[{#1}]{\textup{Appendix~\ref*{#1}}}}
\renewcommand{\eqref}[1]{\hyperref[{#1}]{\textup{(\ref*{#1})}}}

\begin{document}
\preprint{AIP/123-QED}
\title{Three-dimensional Coupled PT-symmetric Electronic Resonators}
\author{Ke Yin}
\orcid{0000-0002-8534-216X}
\affiliation{\mbox{College of Electronics and Information Engineering, Sichuan University, Chengdu 610065, China.}}

\author{Kaihao Tang}
\orcid{0000-0002-2375-6164}
\email[Corresponding author(s):~]{tangkh@scu.edu.cn}
\affiliation{College of Electrical Engineering, Sichuan University, Chengdu 610065, China}

\author{Lu Tan}
\orcid{0009-0006-0190-5954}
\affiliation{\mbox{College of Electronics and Information Engineering, Sichuan University, Chengdu 610065, China.}}

\author{Saddam Ibrahim Dawalbait Bakhat}
\affiliation{\mbox{College of Mechatronics and Control Engineering, Shenzhen University, Shenzhen 518060, China.}}

\author{Tianyu Dong}
\orcid{0000-0003-4816-0073}
\affiliation{School of Electrical Engineering, Xi'an Jiaotong University, Xi'an 710049, China}

\author{\\Huacheng Zhu}
\orcid{0000-0002-1737-870X}
\affiliation{\mbox{College of Electronics and Information Engineering, Sichuan University, Chengdu 610065, China.}}

\author{Yang Yang}
\orcid{0000-0003-0764-1575}
\email[Corresponding author(s):~]{yyang@scu.edu.cn}
\affiliation{\mbox{College of Electronics and Information Engineering, Sichuan University, Chengdu 610065, China.}}

\date{November 1, 2024}

\begin{abstract}
In this article, the non-Hermitian characteristics of three-dimensional PT-symmetric coupled electronic resonators are theoretically analyzed. First, the concept of non-Hermitian PT symmetry is illustrated in the context of electronics using a pair of coupled electronic resonators. Two typical configurations of parallel-coupled PT-symmetric electronic trimers are then analyzed. The results indicate that, for the planar configuration, the system can exhibit two phase transitions as the coupling coefficient or gain-loss parameter changes, different from the linear configuration. By comparing system equations based on coupled-mode theory and circuit theory, it is shown that high dimensionality alone is not a sufficient condition for the existence of a higher-order exceptional point; an approximation condition is also required. A modified exceptional point is proposed, and the approximation conditions for the mean deviation $D$ for the real part of the three eigenfrequencies, satisfying $D \leq 1\%$ and $D \leq 0.1\%$, are discussed, respectively. The theoretical results presented in this paper not only reveal the unique non-Hermitian characteristics of high-dimensional PT-symmetric electronic systems but also offer theoretical support for wireless power transmission and wireless sensing technologies.
\end{abstract}

\maketitle

\section{Introduction} \label{sec:intro}
Over the past few decades, non-Hermitian physics has gradually become one of the major research areas in both theoretical and applied physics. Its proposal originates from the mathematical curiosity of whether the Haimiltonian of a quantum system with real energy levels must be Hermitian, \emph{$\mathcal{H} = \mathcal{H}^\dagger$}\cite{bender2016pt}. In turns out that a special class of non-Hermitian system satisfying parity-time (PT) symmetry can also exhibit real eigenvalues under specific parameter conditions. Based on quantum physics, the concept has garnered significant attention, and due to the intriguing features, it has been extensively studied in various other fields, such as optics \cite{guo2009observation,ruter2010observation}, electronics \cite{schindler2011experimental,schindler2012symmetric,assawaworrarit2017robust,choi2018observation,yang2021ultrarobust,yin2022wireless,hao2023frequency,yin2023high}, microwave \cite{bittner2012pt,yu2020phase}, acoustics \cite{zhu2014pt,fleury2015invisible,fleury2016parity,gao2024controlling}, and related subjects. The field of electronics has become a significant area for exploring the characteristics of PT symmetry, primarily due to its experimental convenience. Furthermore, it supports the advancement of cutting-edge electronic and electrical systems, such as robust wireless power transmission \cite{assawaworrarit2017robust,assawaworrarit2020robust,hao2023frequency} and enhanced wireless sensing technologies \cite{yang2021ultrarobust,yin2022wireless,yin2023high}.

Systems with high-dimensional topology have become a major research direction since the proposal of PT symmetry, focusing mainly on novel topological phenomena \cite{okuma2023non,gao2024controlling,simonyan2024non}, high-order exceptional points \cite{schnabel2017pt,heiss2016model,sakhdari2019experimental,sakhdari2022generalized,yin2023high}, enhanced wireless power transmission \cite{wu2022generalized,sakhdari2020robust,hao2023frequency}, \emph{etc}. High-dimensional topology not only provides more design freedom for topological arrangements, but also introduces diverse phenomena with more potential applications. For instance, high-dimensional PT-symmetric systems can exhibit higher-order exceptional points (EPs) \cite{hodaei2017enhanced,yin2023high}, enabling enhanced sensitivity compared to second-order EPs. In circuit implementations, high-dimensional PT-symmetric systems can be constructed by arranging multiple $RLC$ resonators in different ways. The PT symmetry of a specific arrangement of high-dimensional $RLC$ coupled resonator circuits is determined by whether the Hamiltonian describing the system commutes with the PT operator. Revealing the non-Hermitian characteristics of three-dimensional PT-symmetric circuits is crucial for the development of PT symmetry towards higher-order topology. Furthermore, its unique characteristics may provide requisite theoretical support for both multi-transmitter-multi-receiver wireless power transmission \cite{hao2023frequency} and wireless sensing systems with enhanced sensitivity.

In this article, we first give an intuitive interpretation of non-Hermitian PT symmetry from the perspective of electronic utilizing a pair of coupled $RLC$ resonators. Then, the characteristics of three-dimensional PT-symmetric circuits are theoretically studied on the basis of circuit theory. Whether higher-dimensional PT-symmetric circuits possess other novel characteristics will be investigated in future studies. Similar to the circuit topology discussed in Ref.~\onlinecite{schindler2012symmetric}, we consider the parallel topology of three-dimensional PT-symmetric circuits. The system is composed of a $-RLC$ resonator with negative resistance, an $RLC$ resonator with the same amount of positive resistance, and a lossless relay $LC$ resonator, coupled by mutual inductance. Two different coupling configurations are analyzed, \emph{i.e.} linear coupling and planar coupling, respectively. Finally, the conditions for the existence of high-order EPs are discussed based on a comparison of circuit theory and coupled-mode theory.

\section{A Simple Electronic Binary Circuit}
\begin{figure}[!ht]
    \centering
    \includegraphics[width=3.2in]{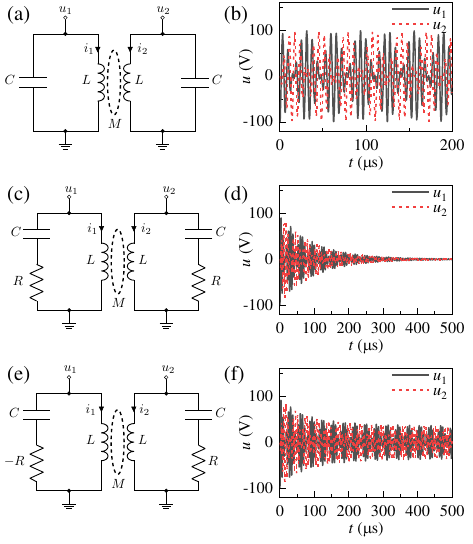}
    \caption{(a) Lossless coupled $LC$ resonator circuit and (b) its zero-state response under given Initial condition.(c) Lossy coupled $LC$ resonator circuit and (d) its zero-state response under given Initial condition.(e) Lossy coupled $LC$ resonator circuit and (f) its zero-state response under given initial condition.}
    \label{fig:figure01}
\end{figure}

Consider a pair of coupled $LC$ resonators shown in \figref{fig:figure01}(a), with circuit parameters $C = 1~\si{nF}$, $L = 1~\si{mH}$, and mutual inductance $M = 0.2~\si{mH}$. In the ideal case where all components are lossless, the system does not exchange energy with the external environment, making it a \textit{closed system}, \emph{i.e.}, the system is Hermitian. When given initial conditions (e.g., $i_1(t=0) = 1~\si{mA}$), the energy in the system propagates without losses between the inductors and capacitors. The voltage signals on both sides of the system contain two frequency components as shown in \figref{fig:figure01}(b), in which case the two eigenfrequencies of the system are both real numbers.

However, in a practical electronic circuit, both inductor and capacitor components have losses, leading to energy exchange between the system and the external environment. In this case, the system becomes an \emph{open system} and exhibits non-Hermitian characteristics. Due to the presence of losses, when given the initial condition $i_1(t = 0) = 1~\si{mA}$, in the absence of an excitation source, the energy in the system will be dissipated by the resistive elements and tend towards zero. Assume that the overall losses in the circuit can be represented by series resistors $R = 20~\si{\ohm}$, as shown in \figref{fig:figure01}(c). In this situation, the voltage on both sides of the system decays exponentially as shown in \figref{fig:figure01}(d), indicating that the eigenfrequencies become complex numbers. 

Notably, there exists a special class of non-Hermitian systems known as \textit{PT-symmetric systems}, with a specific case shown in \figref{fig:figure01}(e). In such systems, the coupled $RLC$ resonators exhibit negative and positive resistances with the same absolute value, yielding balanced gain and loss. When the coupling between the resonators reaches a certain level, the energy generated by the gain resonator is precisely balanced by the energy dissipated by the loss resonator. As a result, the system can achieve a dynamic equilibrium and stably operate at two real eigenfrequencies, as can be seen in \figref{fig:figure01}(f). Therefore, PT-symmetric open systems, under certain parameter conditions, exhibit properties similar to closed systems, which is a counterintuitive characteristic.

\section{Three-Dimensional PT-Symmetric Circuit with Linear-Chain Configuration} \label{sec:linear}
\begin{figure}[!ht]
    \centering
    \includegraphics[width=3.2in]{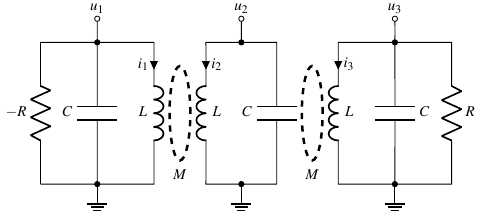}
    \caption{Circuit diagram of a three-dimensional PT-symmetric chainly coupled $LC$ resonator.}
    \label{fig:figure02}
\end{figure}

This section focuses only on the analysis of the parallel topology with mutual inductance coupling. \figref{fig:figure02} shows the circuit schematic of a typical linearly coupled three-dimensional PT-symmetric circuit with a parallel topology. The resonators at both ends have equal values of positive and negative resistances, while the relay resonator is neutral.

According to Kirchoff's current law (KCL) and voltage law (KVL), the capacitor voltages $u_1$, $u_2$, and $u_3$ of the three resonators satisfy the following set of three coupled second-order ordinary differential equations:
\begin{subequations} \label{eq:system_equation_linear}
    \begin{align}
        \frac{\text{d}^2 u_1}{\text{d} \tau^2} - \gamma\frac{\text{d} u_1}{\text{d} \tau} + \frac{1-\kappa^2}{1-2\kappa^2}u_1 &= \frac{\kappa}{1-2\kappa^2}u_2 - \frac{\kappa^2}{1-2\kappa^2}u_3, \\
	    \frac{\text{d}^2 u_2}{\text{d} \tau^2} + \frac{1}{1-2\kappa^2}u_2 &= \frac{\kappa}{1-2\kappa^2}(u_1+u_3), \\
        \frac{\text{d}^2 u_3}{\text{d} \tau^2} + \gamma\frac{\text{d} u_3}{\text{d} \tau} + \frac{1-\kappa^2}{1-2\kappa^2}u_3 &= \frac{\kappa}{1-2\kappa^2}u_2 - \frac{\kappa^2}{1-2\kappa^2}u_1,
    \end{align}
\end{subequations}
where $u_1$, $u_2$, and $u_3$ represent the voltages of the gain, relay, and loss sides, respectively; $\kappa = M/L$ represents the mutual inductance coupling parameter; $\gamma = R^{-1}\sqrt{L/C}$ represents the gain-loss parameter; and $\tau = \omega_0 t = t/\sqrt{LC}$ represents the normalized time variable. Defined state variables as $\bm{\Phi} = (u_1, u_2, u_3, u_1', u_2', u_3')^\text{T}$, the equation \eqref{eq:system_equation_linear} can be reduced to a sate equation with Liouvillian form, which can be expressed as follows: 
\begin{equation} \label{eq:state_equation_linear}
    \frac{\text{d} \bm{\Phi}}{\text{d} \tau} = \mathcal{L} \bm{\Phi},
\end{equation}
where the Liouvillian $\mathcal{L}$ is
\begin{equation} \label{eq:liouville_pt3_linear}
    \mathcal{L} =
        \begin{pmatrix}
         0        & 0          & 0          & 1          & 0        & 0     \\
         0        & 0          & 0          & 0          & 1        & 0     \\ 
         0        & 0          & 0          & 0          & 0        & 1     \\
        -\cfrac{1-\kappa^2}{1-2\kappa^2}  & \cfrac{\kappa}{1-2\kappa^2}   & -\cfrac{\kappa^2}{1-2\kappa^2}    &  \gamma  & 0 & 0   \\
        \cfrac{\kappa}{1-2\kappa^2}       & -\cfrac{1}{1-2\kappa^2}       &  \cfrac{\kappa}{1-2\kappa^2}      &  0        & 0 & 0    \\
        -\cfrac{\kappa^2}{1-2\kappa^2}    & \cfrac{\kappa}{1-2\kappa^2}   & -\cfrac{1-\kappa^2}{1-2\kappa^2}  &  0        & 0 & -\gamma  \\
        \end{pmatrix}.
\end{equation} 

The effective Hamiltonian of the system is given by $\mathcal{H}_\text{eff} = \text{i} \mathcal{L}$. The operators $\mathcal{P}$ and $\mathcal{T}$ are respectively defined as
\begin{subequations} \label{eq:pt_operator}
\begin{align} 
    \mathcal{P} &= \begin{pmatrix}
      \mathbb{J}_3 & 0 \\
      0 & \mathbb{J}_3 \\
    \end{pmatrix}, \\
    \mathcal{T} &= \begin{pmatrix}
      \mathbb{I}_3 & 0 \\
      0 & -\mathbb{I}_3 \\
    \end{pmatrix}\mathscr{K},
\end{align}
\end{subequations}
where $\mathbb{J}_n$ is an $n \times n$ permutation matrix (with $J_{ij} = 1$ when $i+j = n+1$ and $J_{ij} = 0$ otherwise), $\mathbb{I}_n$ is an $n \times n$ identity matrix (with $I_{ij} = 1$ when $i = j$ and $I_{ij} = 0$ otherwise), and $\mathscr{K}$ denotes the complex conjugate transpose. It can be verified that $\mathcal{H}_\text{eff}\mathcal{PT} = \mathcal{PTH}_\text{eff}$, which confirms that the three-dimensional system satisfies PT symmetry. By solving the equation $\det{(\mathcal{L}-\text{i}\nomega\mathbb{I}_6)}=0$ (where $\mathbb{I}_6$ is the six-order identity matrix, $\nomega = \omega/\omega_0$ is the normalized eigenfrequency), the characteristic equation of the system is obtained as
\begin{equation}
    \nomega^6 - \frac{3-\gamma^2-2(1-\gamma^2)\kappa^2}{1-2\kappa^2}\nomega^4 + \frac{3-\gamma^2}{1-2\kappa^2}\nomega^2 - \frac{1}{1-2\kappa^2} = 0.
\label{eq:char_eq}
\end{equation}
Here, $c_1 = -[3 - \gamma^2 - 2 (1-\gamma^2) \kappa^2] c_3$, $c_2 = (3-\gamma^2) c_3$, and $c_3 = 1/(1-2\kappa^2)$ are defined as the coefficients of the quartic, quadratic, and constant terms, respectively.

\subsection{Parametric Region for Unbroken Phase}
We first analyze the parameter space $(\gamma, \kappa)$ for the PT-symmetric phase of the system. Let $\sigma = \nomega^2$. If the curve of the cubic polynomial $f(\sigma) = \sigma^3 + c_2\sigma^2 + c_1\sigma + c_0$ has three intersections with the positive half-axis of $\sigma$ (\emph{i.e.}, if $f(\sigma)=0$ has three distinct positive real roots), then \eqref{eq:liouville_pt3_linear} has six distinct real solutions. At this point, according to $f'(\sigma) = 0$, the extreme values $\sigma_+$ and $\sigma_-$ of $f(\sigma)$ are given by $\sigma_\pm = \left(-c_2 \mp \sqrt{c_2^2 - 3 c_1}\right)/3$. For the curve of $f(\sigma)$ to have three intersections with the positive half-axis, the conditions are:
\begin{subequations} \label{eq:condition_three_real_roots}
\begin{align}
        c_1^2 -3c_2  &> 0,  \\
        \sigma_\pm &> 0,   \\
        f(\sigma_+) &> 0,     \\
        f(\sigma_-) &< 0 ,     \\
        f(0) &< 0.
\end{align}
\end{subequations}
Thus, we can determine the parameter ranges in the $(\gamma, \kappa)$ space corresponding to the PT-symmetric and PT-broken phases of the system, respectively.

\begin{figure}[!ht]
    \centering
    \includegraphics[width=3.0in]{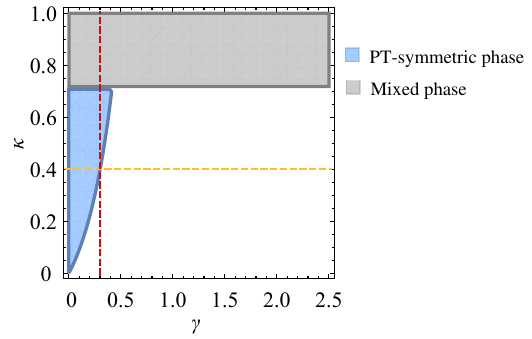}
    \caption{PT-symmetric/broken phase parameter range of three-dimensional PT-symmetric circuit with linear configuration.}
    \label{fig:figure03}
\end{figure}

Figure \ref{fig:figure03} shows the ranges of parameters $\gamma$ and $\kappa$ corresponding to the PT-symmetric and broken phases of the system for $0 \leq \gamma \leq 2.5$ and $0 \leq \kappa \leq 1$. The blue region represents the PT-symmetric phase, while the white region represents the PT-broken phase. Moreover, the PT-symmetric phase exists in the weak gain-loss and strong coupling region since $\gamma$ in parallel circuits denotes the damping rate of the system, while the coupling coefficient signifies the strength of energy exchange between gain and loss resonators. Thus, as $\gamma$ decreases and $\kappa$ increases, the damping decreases and the energy exchange between resonators strengthens, making it easier for the system to reach a stable state, \emph{i.e.}, the PT-symmetric phase. Additionally, due to the restriction of $f(0) < 0$, the system exists only in the symmetric phase when $0<\kappa<\sqrt{2}/2$. It is noteworthy that $\kappa \in (0,\sqrt{2}/2)$ holds physical significance; specifically, $\kappa = \sqrt{2}/2$ represents the maximum value of the coupling parameter when three linearly distributed inductances are fully coupled (\emph{i.e.}, the maximum value of $\kappa$), which can be verified using the law of energy conservation \cite{yin2023high}.

\subsection{Eigenfrequency Evolution} \label{subsec:linear-B}
\begin{figure}[!ht]
    \centering
    \includegraphics[width=3.2in]{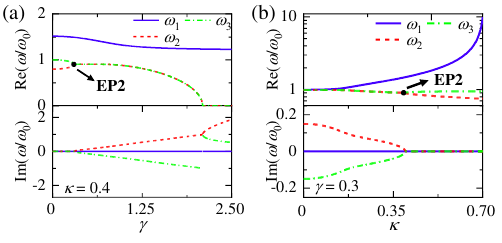}
    \caption{Eigenfrequency evolution of a linear three-level PT-symmetric circuit when (a) $\kappa = 0.4$ and $\gamma$ is changing; (b) $\gamma = 0.3$ and $\kappa$ is changing.}
    \label{fig:figure04}
\end{figure}

By using the Cardano's formula to solve the characteristic equation \eqref{eq:char_eq}, we can obtain the six characteristic frequencies of the system as follows:
\begin{subequations} \label{eq:eigenvalues}
    \begin{align}
        \omega_{1,4} &= \pm \sqrt{s+t-\frac{1}{3}c_1}, \\
        \omega_{2,5} &= \pm \sqrt{-\frac{1}{2}(s+t) + \text{i}\frac{\sqrt{3}}{2}(s-t) - \frac{1}{3}c_1}, \\
        \omega_{3,6} &= \pm \sqrt{-\frac{1}{2}(s+t) - \text{i}\frac{\sqrt{3}}{2}(s-t) - \frac{1}{3}c_1}, 
    \end{align}
\end{subequations}
where
\begin{subequations} \label{eq:cardano_formula}
    \begin{align}
        s &=  \sqrt[3]{p + \sqrt{p^2 + q^3}}, \\
        t &=  \sqrt[3]{p - \sqrt{p^2 + q^3}}, \\
        p &= -\frac{1}{27}c_1^3 + \frac{1}{6}c_1 c_2 - \frac{1}{2}c_3, \\
        q &= -\frac{1}{9}c_1^3 + \frac{1}{3}c_2.
    \end{align}
\end{subequations}
Here, the discriminant is defined as $\Delta = p^2 + q^3$. Consider the variation in the system's eigenfrequencies with respect to $\gamma$ when $\kappa = 0.4$ (corresponding to the yellow dashed line in \figref{fig:figure03}). With $\Delta = 0$, we obtain $\gamma_\ep = 0.3$. When $\gamma < \gamma_\ep$, $\Delta<0$, indicating the existence of three pairs of distinct real roots that are each other's opposites. When $\gamma = \gamma_\ep$, $\Delta = 0$, resulting in a pair of opposite real roots and two pairs of identical real roots that are opposites. When $\gamma > \gamma_\ep$, $\Delta > 0$, leading to a pair of opposite real roots and two pairs of conjugate complex roots with opposite real parts. \figref{fig:figure04}(a) illustrates the evolution of three eigenfrequencies having a positive real part with respect to $\gamma$. It can be observed that the system is in PT-symmetric phase when $\gamma<0.3$, and in a broken phase when $\gamma > 0.3$. Considering a fixed $\gamma$, the behavior of the eigenfrequencies concerning changes in $\kappa$ is examined (corresponding to the red dashed line in \figref{fig:figure03}). Let $\gamma=0.3$, and from $\Delta = 0$, we derive $\kappa_\ep = 0.397$. Similarly, \figref{fig:figure04}(b) demonstrates the evolution of real and imaginary parts of the three eigenfrequencies with positive real parts. It is evident that when $\kappa < \kappa_\ep$, the eigenfrequencies consist of one real number $\omega_1$ and a pair of complex conjugates $\omega_{2,3}$. When $\kappa > \kappa_\ep$, the eigenfrequencies comprise three distinct real numbers. It is worth noting that in the aforementioned cases, the frequency evolution of the three-dimensional PT-symmetric circuit does not exhibit third-order EPs similar to those of optical systems \cite{schnabel2017pt,heiss2016model}. The spontaneous symmetry breaking phase transition point occurs at a second-order EP, where only two modes corresponding to $\omega_2$ and $\omega_3$ become degenerate. This phenomenon will be discussed further in \secref{sec:hoep}.

\subsection{Eigenvector Evolution} \label{subsec:linear-C}
The eigenfrequencies provide information about the operating frequencies of the circuit, while the eigenvectors provide information about the amplitude and phase. Next, we analyze the evolution of the eigenvectors in the parameter space. The eigenvalue equation corresponding to the state equation in \eqref{eq:state_equation_linear} is given by $\mathcal{L} \bm{\Phi}_n = \inomega_n \bm{\Phi}_n$ or $\mathcal{H}_\text{eff}\bm{\Phi}_n = \nomega_n \bm{\Phi}_n$, where $\bm{\Phi}_n$ represents the eigenvector for $\nomega_n$. The components $(u_1, u_2, u_3)^\text{T}$ of the eigenvector can be derived as
\begin{equation}\label{eq:eigenvector_pt3_linear}
    \begin{pmatrix}
        u_1   \\
        u_2   \\
        u_3   
    \end{pmatrix}=
    \begin{pmatrix}
        1   \\
        \cfrac{2\kappa(1-\nomega_n^2)}{(1-\nomega_n^2 + \im \gamma \nomega_n)[1-(1-2\kappa^2) \nomega_n^2]} \\[1em]
        -\cfrac{\gamma \nomega_n + \im(1-\nomega_n^2)}{\gamma \nomega_n - \im (1-\nomega_n^2)}
    \end{pmatrix}.
\end{equation}
Correspondingly, the components $(u_1', u_2', u_3')^\text{T}$ are represented as $(u_1', u_2', u_3')^\text{T} = \inomega_n (u_1, u_2, u_3)^\text{T}$. In the PT-symmetric phase where $\nomega_n$ is real, \eqref{eq:eigenvector_pt3_linear} can be rewritten in exponential form as
\begin{equation}\label{eq:eigenvector_pt3_linear_exp}
    \begin{pmatrix}
        u_1& 
        u_2&
        u_3   
    \end{pmatrix}^\text{T}=
    \begin{pmatrix}
        1 &
        A_n \Exp^{\im \alpha_n} &
       \Exp^{\im \beta_n} 
    \end{pmatrix}^\text{T},
\end{equation}
where $A_n = \left|\frac{2\kappa(1-\nomega_n^2)}{\sqrt{\gamma^2\nomega_n^2+(1-\nomega_n^2)^2}[1-(1-2\kappa^2)\nomega_n^2]}\right|$, $\alpha_n = -\tan ^{-1}[(1-\nomega_n^2)(\gamma \nomega_n]$, $\beta_n = -2\alpha_n$, and $n = \{1,2,3\}$. From \eqref{eq:eigenvector_pt3_linear_exp}, it can be observed that in the PT-symmetric phase, the oscillation amplitudes of $u_1$ on the gain side and $u_3$ on the loss side are equal, with a phase difference of $\beta$; the difference in oscillation amplitudes between $u_1$ on the gain side and $u_2$ in the relay is $A_n$, with a phase difference of $\alpha$.

\begin{figure}[!ht]
    \centering
    \includegraphics[width=3.2in]{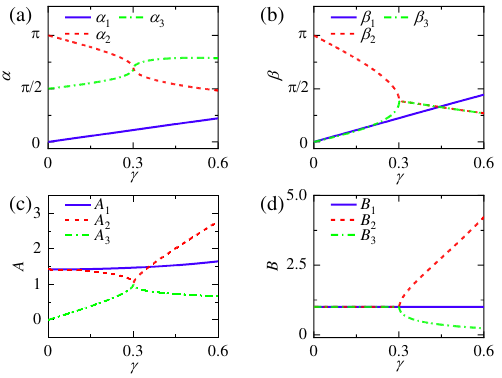}
    \caption{(a), (b)The evolution of phase differences $\alpha$ and $\beta$ of each mode as a function of $\gamma$. (c), (d)The evolution of amplitude ratios of each mode as a function of $\gamma$. Here, $\kappa_1 = 0.1$ and $\kappa_2 = 0.4$.}
    \label{fig:figure05}
\end{figure}

Figures \ref{fig:figure05}(a) and \ref{fig:figure05}(b) respectively depict the phase differences $\alpha_n$ and $\beta_n$ of each eigenfrequency $\nomega_n$ with $\kappa=0.4$ against the gain-loss parameter $\gamma$. As shown in \figref{fig:figure05}(a), when $\gamma=0$, the phase differences between the relay resonator and the gain resonator at three frequencies are $0$, $\uppi/2$, and $\uppi$, respectively. With increasing $\gamma$, $\nomega_2$ and $\nomega_3$ gradually approach each other until degeneracy occurs at the EP; their phase differences also gradually approach equality. Simultaneously, the eigenvalues $\nomega_{2,3}$ corresponding to the eigenmodes $\Phi_{2,3}$ also degenerate at the EP. With an increase of $\gamma$, the phase difference for $\nomega_1$ increases. \figref{fig:figure05}(b) illustrates the voltage phase differences between the loss- and gain-side resonators concerning $\gamma$. When $\gamma = 0$, the phase differences corresponding to the modes $\nomega_{1,3}$ are zero, and for $\nomega_2$, it is $\uppi$.

As $\gamma$ increases, the phase differences of $\nomega_{2,3}$ gradually approach each other and become degenerate at the EP, while the phase difference for $\nomega_1$ linearly increases. The analysis also presents the variation in the amplitude of the $u_2$ component of each eigenvector in the symmetric phase. It is essential to note that this representation is not the actual amplitude ratio between $u_2$ and $u_1$; the actual ratio also depends on the initial conditions of the circuit, which will be discussed in subsequent sections. \figref{fig:figure05}(c) and \figref{fig:figure05}(d) display the amplitude variations of the eigenvectors with respect to $\gamma$ when $\kappa = 0.4$. It is observed that the amplitude ratios between the relay side and the gain-loss side resonators are no longer equal to $1$. Additionally, with increasing $\gamma$, the amplitudes of the component $u_2$ in the eigenvectors $\Phi_2$ and $\Phi_3$ gradually approach each other and become degenerate at the EP, corresponding to the degeneracy of $\nomega_{2,3}$ as shown in \figref{fig:figure03}.

\subsection{Dynamic Properties} \label{subsec:linear-D}
By solving the coupled second-order ordinary differential equations, the time-domain voltage waveforms for various parameters within the system can be obtained. Here, we treat the linear homogeneous equation \eqref{eq:state_equation_linear} based on the Jordan-Chevalley decomposition \cite{hsieh1999basic}. With the given parameter $\gamma$ and $\kappa$, all the eigenvalues of the system can be calculated by \eqref{eq:eigenvalues}. Let $\nomega_j$ ($j = 1, 2, ..., k$) be the distinct eigenvalues of $\mathcal{L}$ and let $m_j$ ($j = 1, 2, \ldots, k$) be their respective multiplicities. From the analysis in \secref{subsec:linear-C}, it is known that the system exhibits six distinct eigenvalues in the PT-symmetric and PT-broken phase, while exhibiting four distinct eigenvalues at the phase transition point (2nd-order EP). Then, the characteristic polynomial of the matrix $\mathcal{L}$ is given by $p(\nomega) = (\nomega - \inomega_1)^{m_1}(\nomega - \inomega_2)^{m_2} \cdots (\nomega - \inomega_k)^{m_k}$. Decompose $\displaystyle \frac{1}{p(\nomega)}$ into partial fractions $\displaystyle \frac{1}{p(\nomega)} = \sum_{j=1}^{k} \frac{q_j(\nomega)}{(\nomega-\inomega_j)^{m_j}}$, where $q_j(\nomega)$ is the denominator polynomial of the $j$-th term resulting in $\sum_{j=1}^k q_j(\nomega)\prod_{h \neq j}(\nomega-\inomega_h)^{m_h} = 1$. Then projection polynomial $p_j(\nomega) = q_j(\nomega) \prod_{h \neq j} (\nomega-\inomega_h)^{m_h}$ and we can obtain the projection matrix $ \bm{P}_j(\mathcal{L})$ as
\begin{equation}
    \bm{P}_j(\mathcal{L}) = \bm{Q}_j(\mathcal{L})\prod_{h \neq j}(\mathcal{L}-\inomega_h \mathbb{I}_n)^{m_h}.
\end{equation}

Decompose $\mathcal{L}$ into $\mathcal{L} = \bm{S} + \bm{N}$, where $\bm{S}$ and $\bm{N}$ are respectively obtained by 
\begin{subequations}
    \begin{align}
        \bm{S} &= \sum_{j=1}^k \inomega_j \bm{P}_j, \\
        \bm{N} &= \mathcal{L} - \bm{S}.
    \end{align}
\end{subequations}
A fundamental matrix solution $\Exp^{\tau \mathcal{L}}$ can then be calculated based on the above S-N decomposition, which is
\begin{equation}
    \Exp^{\tau \mathcal{L}} = \sum_{j=1}^{k}e^{\inomega_j \tau}\left[ \mathbb{I}_6 + \sum_{h=1}^{5}\frac{\tau^h}{h!} \bm{N}^h \right] \bm{P}_j(\mathcal{L}).
\end{equation}
With the initial conditions $\bm{\Phi}(\tau=0)$, the state variables can be obtained by
\begin{equation} \label{eq:state_variable_pt3}
    \begin{split}
        \bm{\Phi} &= \Exp^{\tau \mathcal{L}} \bm{\Phi}(\tau=0),
    \end{split}
\end{equation}

\begin{figure}[!ht]
    \centering
    \includegraphics[width=3.2in]{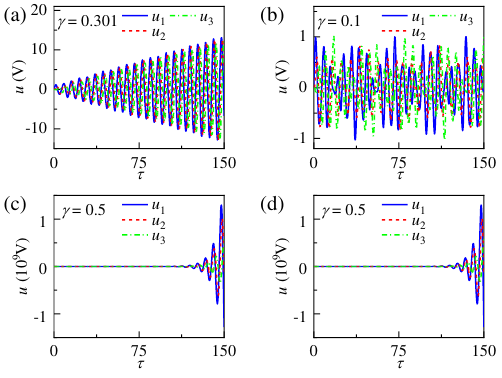}
    \caption{Time-domain voltage waveform when (a) $\gamma = 0.1$, (b) $\gamma = \gamma_\ep = 0.301$, (c) $\gamma = 0.5$ and (d) $\gamma = 2$. Here, $\kappa = 0.4$.}
    \label{fig:figure06}
\end{figure}

Figure \ref{fig:figure06} displays the time-domain responses of the resonators under different parameters when $\kappa = 0.4$, with the initial conditions $\bm{\Phi}(\tau=0) = (0, 0, 0, 1, 0, 0)^\text{T}$. In the symmetric phase, all resonators exhibit equal-amplitude oscillations with three frequency components. Specifically, the $RLC$ resonators on the gain and loss sides show equal amplitudes but different phases. However, the $LC$ resonator on the relay side has a slightly smaller temporal voltage waveform amplitude compared to the gain and loss side voltages, leading to $|u_2/u_1| = 0.815$ derived from \eqref{eq:state_variable_pt3}, as depicted in \figref{fig:figure06}(a). At the phase-transition point, the system enters an unstable working state. The amplitude shows linear growth, representing a critical state between symmetric equal-amplitude oscillations and exponential growth in the broken phase, as shown in \figref{fig:figure06}(b). In the broken phase, the voltage amplitudes of all three resonators exponentially grow, indicating an unstable state. When the value of $\gamma$ is small, the system exhibits an oscillatory growth, corresponding to the underdamped broken phase (Broken Phase I). However, when the $\gamma$ value is large, it demonstrates non-oscillatory growth, depicting the overdamped broken phase (Broken Phase II), as illustrated in \figref{fig:figure06}(c) and \figref{fig:figure06}(d).

\section{Three-Dimensional PT-Symmetric Circuit with Planar Configuration}
\begin{figure}[!ht]
    \centering
    \includegraphics[width=2.2in]{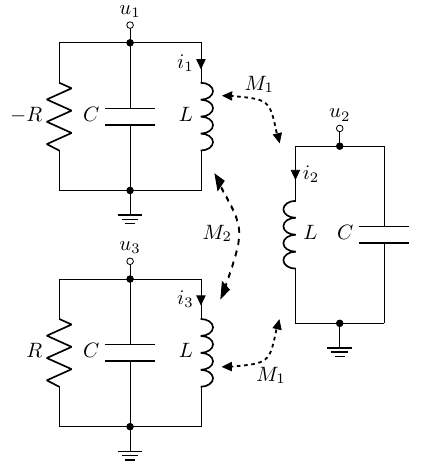}
    \caption{Circuit diagram of a three-dimensional coupled PT-symmetric circuit with planar configuration.}
    \label{fig:figure07}
\end{figure}

The schematic diagram of a three-dimensional PT-symmetric coupled $LC$ resonator circuit with a planar structure is shown in \figref{fig:figure07}. In this case, the three resonators are coupled to each other in pairs. To ensure symmetry of the circuit topology, it is necessary for the coupling parameters between the relay resonator and the gain resonator, as well as between the relay resonator and the loss resonator, to be equal.

The system equations of the circuit can be derived from Kirchhoff's laws, resulting in the coupled second-order ordinary differential equations given by
\begin{subequations} \label{eq:system_equation_pt3_planar}
    \begin{align}
        \frac{\text{d}^2 u_1}{\text{d} \tau^2} - \gamma \frac{\text{d} u_1}{\text{d} \tau} + \frac{1-\kappa_1^2}{(1-\kappa_2)\xi}u_1 &= \frac{\kappa_1}{\xi}u_2 + \frac{\kappa_2-\kappa_1^2}{(1-\kappa_2)\xi}u_3, \\
        \frac{\text{d}^2 u_2}{\text{d} \tau^2} + \frac{1+\kappa_2}{\xi}u_2 &= \frac{\kappa_1}{\xi}(u_1+u_3), \\
        \frac{\text{d}^2 u_3}{\text{d} \tau^2} + \gamma \frac{\text{d} u_3}{\text{d} \tau} + \frac{1-\kappa_1^2}{(1-\kappa_2)\xi}u_3 &= \frac{\kappa_1}{\xi}u_2 + \frac{\kappa_2-\kappa_1^2}{(1-\kappa_2)\xi}u_1,
    \end{align}
\end{subequations}
where $\xi = 1 + \kappa_2 - 2 \kappa_1^2$. Similarly to \eqref{eq:state_equation_linear}, the coefficient matrix $\mathcal{L}$ of the state equation is
\begin{equation} \label{eq:liouv_linear}
    \mathcal{L} =
    \begin{pmatrix}
      0        & 0          & 0          & 1          & 0        & 0     \\
      0        & 0          & 0          & 0          & 1        & 0     \\   
      0        & 0          & 0          & 0          & 0        & 1    \\
      -\cfrac{1-\kappa_1^2}{(1-\kappa_2)\xi}         &  \cfrac{\kappa_1}{\xi}     &  \cfrac{\kappa_2-\kappa_1^2}{(1-\kappa_2)\xi}   &  \gamma   & 0 & 0   \\[1em]
      \cfrac{\kappa_1}{\xi}                          & -\cfrac{1+\kappa_2}{\xi}   &  \cfrac{\kappa_1}{\xi}                          &  0        & 0 & 0    \\[1em]
      \cfrac{\kappa_2-\kappa_1^2}{(1-\kappa_2)\xi}   & \cfrac{\kappa_1}{\xi}      & -\cfrac{1-\kappa_1^2}{(1-\kappa_2)\xi}          &  0        & 0 & -\gamma  \\
    \end{pmatrix}.
\end{equation} 

By verifying $\mathcal{H}_\text{eff}\mathcal{PT}= \mathcal{PTH}_\text{eff}$ (where $\mathcal{H}_\text{eff}=\text{i}\mathcal{L}$), we can confirm that this three-dimensional system satisfies PT symmetry. From the coefficient matrix, we can derive the characteristic equation of the system as $\det{(\mathcal{L}-\text{i}\nomega\mathbb{I}_6)}=0$, which can be written as:
\begin{equation}
\begin{split}
    \nomega^6 &+ \frac{\gamma^2(1-\kappa_2)\xi+2 \kappa_1^2+\kappa_2^2-3}{(1-\kappa_2)\xi}\nomega^4 \\
    &+ \frac{3-\gamma^2(1-\kappa_2^2)}{(1-\kappa_2)\xi}\nomega^2 - \frac{1}{(1-\kappa_2)\xi} = 0.
\end{split}  
\end{equation}

\subsection{Parametric Region for Unbroken Phase}
\begin{figure}[!ht]
    \centering
    \includegraphics[width=3.0in]{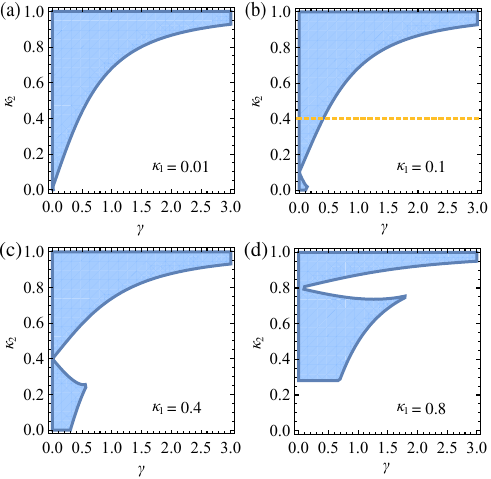}
    \caption{The parametric region of PT-symmetric phase in the parameter space $(\gamma, \kappa_2)$ when (a) $\kappa_1 = 0.01$, (b) $\kappa_1 = 0.1$, (c) $\kappa_1 = 0.4$, and (d) $\kappa_1 = 0.8$.}
    \label{fig:figure08}
\end{figure}

According to the conditions given by \eqref{eq:condition_three_real_roots}, the parameter range for the PT-symmetric phase in the parameter space can be obtained. Unlike the case of chain coupling, the system's parameter space is three-dimensional, \emph{i.e.}, $(\gamma, \kappa_1, \kappa_2)$. By fixing the value of $\kappa_1$, we can determine the range of $(\gamma, \kappa_2)$ that satisfies the symmetric phase, as shown by the blue region in \figref{fig:figure08}. As discussed earlier, in a chain structure, the system only undergoes one phase transition as a certain parameter changes. However, from \figref{fig:figure08}, we can see that when $\kappa_2$ changes while $\gamma$ is fixed, the system can exhibit two ranges of PT-symmetric phase, indicating a transition process from the PT-symmetric phase to the PT-broken phase and back to the PT-symmetric phase. On the other hand, when $\kappa_2$ changes while $\gamma$ is fixed, there is only one phase transition process.

\begin{figure}[!ht]
    \centering
    \includegraphics[width=3.0in]{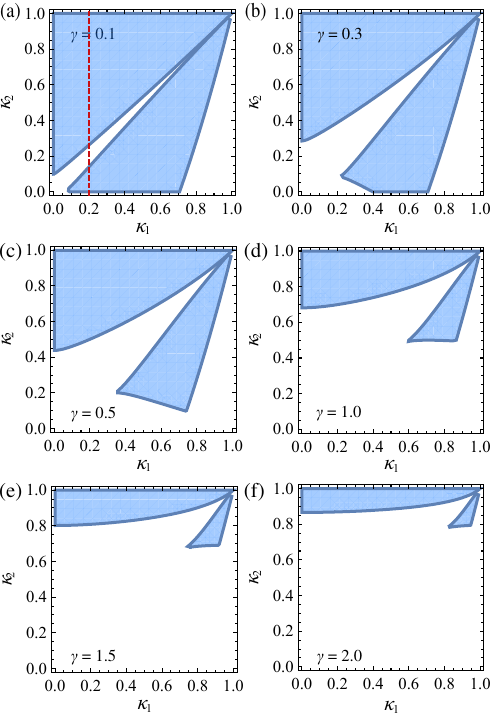}
    \caption{The parametric region of PT-symmetric phase in the parameter space $(\kappa_1,\kappa_2)$ when (a) $\gamma = 0.1$, (b) $\gamma = 0.3$, (c) $\gamma = 0.5$, (d) $\gamma = 1.0$, (e) $\gamma = 1.5$, and (f) $\gamma = 2.0$.}
    \label{fig:figure09}
\end{figure}

Similarly, we can analyze the range of $(\kappa_1, \kappa_2)$ that satisfies the PT-symmetric phase for different values of $\gamma$, as shown by the blue region in \figref{fig:figure09}. It can be observed that as $\gamma$ increases, the range of the PT-symmetric phase gradually decreases and is mainly distributed in the strong coupling region, \emph{i.e.}, the region where $\kappa_2$ has larger values. This indicates that the degree of energy exchange in the system is mainly related to the magnitude of coupling between the gain and loss resonators. Additionally, when one of the parameters, either $\kappa_1$ or $\kappa_2$, is fixed, the system can undergo two transition processes as the other parameter changes.

\subsection{Eigenfrequency Evolution}
\begin{figure}[!ht]
    \centering
    \includegraphics[width=3.2in]{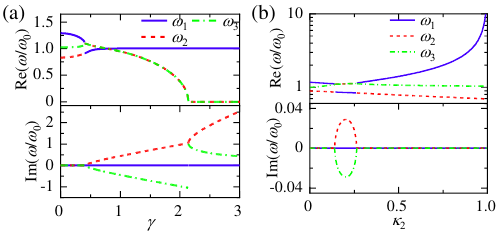}
    \caption{(a) Eigenfrequency evolution as a function of $\gamma$ when $\kappa_1 = 0.1,\kappa_2 = 0.4$. (b) Eigenfrequency evolution as a function of $\kappa_2$ when $\kappa_1 = 0.2$ and $\gamma = 0.1$.}
    \label{fig:figure10}
\end{figure}

After determining the ranges of PT-symmetric and PT-broken phases in the parameter space, further discussion regarding the evolution of eigenfrequencies during the system's phase transition for different given parameters can be initiated. Using the same analysis method as in \secref{subsec:linear-B}, three sets of eigenfrequencies that are each other's opposites in a planar coupled system can be obtained. \figref{fig:figure10}(a) illustrates the evolution of the real and imaginary parts of the three positive real eigenfrequencies of the system when the coupling parameters $\kappa_1$ and $\kappa_2$ are fixed, with respect to $\gamma$ (corresponding to the yellow dashed line in \figref{fig:figure08}). It can be observed that when $\gamma < \gamma_\ep$, all three eigenfrequencies are real numbers; when $\gamma_\ep < \gamma < \gamma_\text{C}$, the eigenfrequencies consist of one real number and a pair of complex conjugates; when $\gamma > \gamma_\text{C}$, the eigenfrequencies comprise one real number and two purely imaginary numbers. Therefore, the parameter range for PT-symmetric phase is $\gamma \in (0, \gamma_\ep)$; symmetry breaks when $\gamma > \gamma_\ep$. Similarly, at this point, the system only exhibits a second-order singularity, indicating that only two modes degenerate at the phase transition point. \figref{fig:figure10}(b) displays the evolution of the real and imaginary parts of the three positive real eigenfrequencies of the system when the coupling parameters $\kappa_1$ and $\gamma$ are fixed, with respect to $\kappa_2$ (corresponding to the red dashed line in \figref{fig:figure09}). Here, two phase transition processes occur from PT-symmetric to broken and back to PT-symmetric phases: when $\kappa_2 < \kappa_\text{C1}$, the eigenfrequencies comprise three distinct real numbers, indicating a PT-symmetric phase; when $\kappa_\text{C1}<\kappa_2 < \kappa_\text{C2}$, the eigenfrequencies consist of one real number and a pair of complex conjugates, signifying symmetry breaking; when $\kappa_2 > \kappa_\text{C2}$, the eigenfrequencies again become three distinct real numbers, returning to the PT-symmetric phase. Here, $\kappa_\text{C1}$ and $\kappa_\text{C2}$ are both second-order EPs, corresponding to the degeneracy of modes $\omega_1$ and $\omega_3$. Additionally, since $\kappa_2 = 1$ is a divergent singularity point where the corresponding eigenfrequency $\omega_1 \to \infty$, the value of eigenfrequency $\omega_1$ sharply increases as $\kappa_2 \to 1$. When $\kappa_1$ is varied while keeping $\kappa_2$ and $\gamma$ constant, the system undergoes a phase transition similar to \figref{fig:figure10}, and the evolution of eigenfrequencies follows a similar pattern, which is not reiterated here. It is noteworthy that when a specific value is assigned to $\kappa_2$, the upper limit of $\kappa_1$ is no longer 1 but is determined by the solution to $1 + \kappa_2 - 2 \kappa_1^2 = 0$, which corresponds to the divergent singularity point for $\kappa_1$. This conclusion can also be derived from energy conservation principles.

\subsection{Eigenvector Evolution}
Further discussion focuses on the evolution of the eigenvectors in parameter space. Based on the eigenvalue equation $\mathcal{L} \bm{\Phi}_n = \inomega_n \bm{\Phi}_n$ or $\mathcal{H}_\text{eff} \bm{\Phi}_n = \nomega_n \bm{\Phi}_n$, where $\bm{\Phi}_n$ represents the eigenvector corresponding to $\nomega_n$, the components $(u_1,u_2,u_3)^\text{T}$ in the eigenvector can be calculated as
\begin{equation}
    \begin{pmatrix}
        u_1   \\
        u_2   \\
        u_3   
    \end{pmatrix}=
    \begin{pmatrix}
    1   \\
    - \cfrac{2 \kappa_1 [1-(1-\kappa_2)\nomega_n^2]}{(\xi\nomega_n^2-\kappa_2-1)[1-(1-\kappa_2)\nomega_n^2 + \im \gamma \nomega_n (1-\kappa_2)]} \\[1em]
    - \cfrac{\gamma \nomega(1-\kappa_2) + \im(1-\nomega ^2+ \kappa_2 \nomega_n^2)}{\gamma \nomega(1-\kappa_2) - \im(1-\nomega ^2+ \kappa_2 \nomega_n^2)}
    \end{pmatrix}.
\end{equation}
Correspondingly, the components $(u_1', u_2', u_3')^\text{T}$ can be expressed as $(u_1', u_2', u_3')^\text{T} = \inomega_n(u_1,u_2,u_3)^\text{T}$. In the PT-symmetric phase, when $\nomega_n$ is real, the exponential form of the eigenvectors matches \eqref{eq:eigenvector_pt3_linear_exp}, but the amplitudes and phases are no longer identical. In this scenario, we have
\begin{subequations}
\begin{align}
A_n      &= \left|\frac{2 \kappa_1 (\kappa _2 \nomega_n^2-\nomega_n^2+1)}{(\xi \nomega_n^2-\kappa_2-1) \sqrt{\gamma^2\nomega_n^2(1-\kappa_2)^2+(\kappa_2 \nomega_n^2-\nomega_n^2+1)^2}}\right|, \\
\alpha_n &= \tan ^{-1}\left[-\frac{\gamma \nomega_n(1-\kappa_2)}{1-(1-\kappa_2) \nomega_n^2}\right], \\
\beta_n  &= 2 \tan^{-1}\left[\frac{1-\nomega_n^2(1-\kappa_2)}{\gamma\nomega_n(1-\kappa_2)}\right],
\end{align}
\end{subequations}
where $n = \{1,2,3\}$. Similar to the case of three-dimensional chained coupling, in the PT-symmetric phase, $u_1$ and $u_3$ have equal oscillation amplitudes with a phase difference of $\beta_n$. $u_1$ and $u_2$ have an amplitude difference of $A_n$ and a phase difference of $\alpha_n$.

\begin{figure}[!ht]
    \centering
    \includegraphics[width=3.2in]{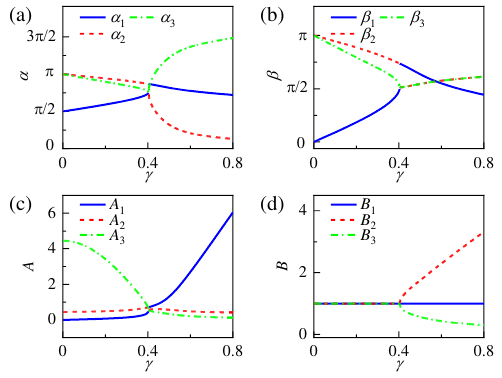}
    \caption{(a), (b) The evolution of phase differences $\alpha$ and $\beta$ of each mode as a function of $\gamma$. (c), (d) The evolution of amplitude ratios of each mode as a function of $\gamma$. Here, $\kappa_1 = 0.1$ and $\kappa_2 = 0.4$.}
    \label{fig:figure11}
\end{figure}

\figref{fig:figure11}(a) illustrates the phase difference $\alpha$ between the relay and gain sides. When $\gamma=0$, the modal phase difference for $\nomega_{2,3}$ is $\uppi$, while for $\nomega_1$, it is $\uppi/2$. As $\gamma$ increases, the phase of modes $\nomega_1$ and $\nomega_3$ gradually approaches and eventually coincides at the EP. For the loss-side resonator, when $\gamma = 0$, the modal phase difference for $\nomega_{2,3}$ is also $\uppi$, while for $\nomega_1$, it is zero. Similar to the gain-relay side, as $\gamma$ increases, the phase of modes $\nomega_1$ and $\nomega_3$ gradually approaches and eventually coincides at the EP, as shown in \figref{fig:figure11}(b). \figref{fig:figure11}(c) and \figref{fig:figure11}(d) respectively depict the amplitude ratios between the $u_{2,3}$ components and the $u_1$ component in the eigenvector. It is evident that in the symmetric phase, the amplitude ratio between the loss side and the gain side voltages remains $|u_3/u_1| = 1$. However, the amplitude ratios between the relay and gain sides vary across modes and are also influenced by the circuit's initial conditions.

\subsection{Dynamic Properties}
\begin{figure}[!ht]
    \centering
    \includegraphics[width=3.2in]{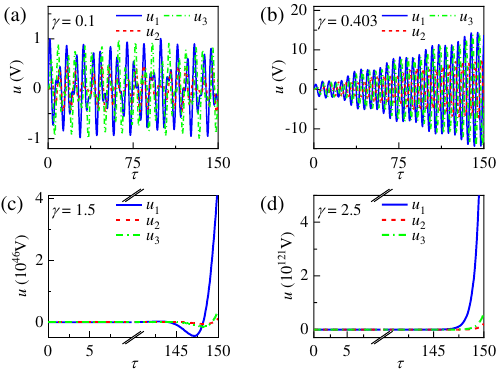}
    \caption{Transient voltage waveform in different parametric phases when $\kappa_1 = 0.2$ and $\kappa_2 = 0.4$. (a) PT-symmetric phase when $\gamma = 0.1$; (b) Phase transition point when $\gamma = \gamma_\text{EP} = 0.403$; (c) PT-broken phase I when $\gamma = 1.5$; (d) PT-broken phase II when $\gamma = 2.5$.}
    \label{fig:figure12}
\end{figure}

Next we analyze the dynamic properties in different parametric region, focusing on the time-domain dynamic characteristics of the system under different parameter phases. The methods for solving the voltage on the gain, relay, and loss sides in the case of planar coupling are the same as described in \secref{subsec:linear-D}. \figref{fig:figure12} presents the time-domain waveforms of $u_1$, $u_2$, and $u_3$ obtained through numerical solutions for $\kappa_1 = 0.1$ and $\kappa_2 = 0.4$ at different $\gamma$ values. The initial conditions chosen here are $\bm{\Phi}(\tau=0) = (0,0,0,1,0,0)^\text{T}$. As depicted in \figref{fig:figure12}(a), when $\gamma = 0.1$, the system is in the PT-symmetric phase. The time-domain voltage waveforms across the three resonators resemble those in \figref{fig:figure05}(c), exhibiting stable amplitude oscillations at three frequency components. The amplitudes of the gain and loss resonators are equal, consistent with the amplitude ratio results in \figref{fig:figure11}(d). However, the relay resonator exhibits a smaller amplitude, yielding a calculated amplitude ratio of $|u_2/u_1| = 0.475$ between the relay and gain sides. At the phase transition point where $\gamma = \gamma_\text{EP} = 0.403$, the system enters an unstable operating state. Unlike \figref{fig:figure06}(b), due to the closeness of the three eigenfrequencies at the transition point, the voltage amplitude does not increase linearly, displaying noticeable amplitude oscillations as shown in \figref{fig:figure12}(b). Upon surpassing the phase transition point as $\gamma$ continues to increase, the system undergoes spontaneous symmetry breaking. The voltages across the three resonators exhibit exponential divergence. As $\gamma$ increases beyond the transition point, the initially underdamped exponential oscillations (as shown in \figref{fig:figure12}(c)) evolve into overdamped exponential growth without oscillations (as depicted in \figref{fig:figure12}(d)).

\begin{figure}[!ht]
    \centering
    \includegraphics[width=3.2in]{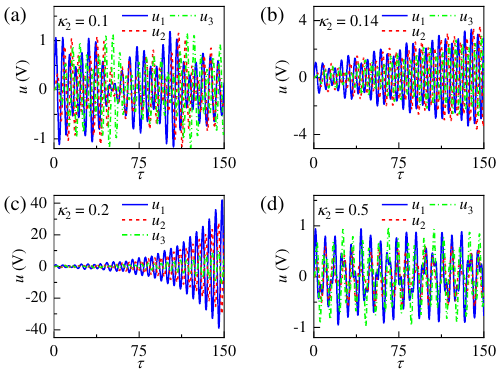}
    \caption{Transient voltage waveform in different parametric phases when $\kappa_1 = 0.2$ and $\gamma = 0.2$. (a) PT-symmetric phase I when $\kappa_2 = 0.1$; (b) First phase transition point when $\kappa_2 = \kappa_\text{EP} = 0.14$; (c) PT-broken phase when $\kappa_2 = 0.2$; (d) PT-symmetric phase II when $\kappa_2 = 0.5$.}
    \label{fig:figure13}
\end{figure}

Figure \ref{fig:figure13} illustrates the variations in the system's time-domain dynamics under different $\kappa_2$ values while maintaining $\kappa_1 = 0.2$ and $\gamma = 0.1$. As shown in \figref{fig:figure13}(a), when $\kappa_2 = 0.1$, the system is in PT-symmetric phase, exhibiting stable amplitude oscillations across three frequency components. Interestingly, unlike the symmetric phase waveform in \figref{fig:figure12}(a), at this point the voltage amplitudes across the three resonators are nearly equal. Calculations yield $|u_3/u_1| = 1$ and $|u_2/u_1| = 0.986$. At this stage, the system's energy is distributed nearly evenly among the three resonators, presenting a potential possibility for highly efficient wireless power transmission systems with multiple loads. At the first phase transition point where $\kappa_2 = \kappa_\text{EP} = 0.14$, the amplitudes of each voltage gradually increase. Here, $u_1$ and $u_2$ exhibit equal oscillation amplitudes, while $u_3$ has a smaller amplitude. With further increase in $\kappa_2$, the system undergoes spontaneous symmetry breaking, showcasing exponential oscillatory growth in the voltage waveform, as depicted in \figref{fig:figure13}(c). As $\kappa_2$ continues to increase, the system re-enters the PT-symmetric phase, where the waveform resembles the PT-symmetric phase waveform in \figref{fig:figure12}(a). Specifically, the amplitude of the relay resonator compared to the gain and loss resonators is smaller, with an amplitude ratio of $0.621$, while the gain and loss sides exhibit equal amplitudes, as shown in \figref{fig:figure13}(d).

From \figref{fig:figure13}, it is evident that the dynamic behavior of the three-dimensional PT symmetric circuit with planar coupling is more diverse. Particularly, when the coupling $\kappa_1$ between the gain-loss resonator and the coupling $\kappa_2$ between the gain-relay (loss-relay) resonators is comparable, the amplitude of the relay resonator might equal or even surpass the amplitudes of the gain and loss resonators. This suggests that the three-dimensional PT symmetric circuit with planar coupling could function as a step-up transformer from gain to relay.

\section{Conditions for HOEP in Higher-dimensional PT-smmetric Circuits}\label{sec:hoep}
\begin{figure}[!ht]
    \centering
    \includegraphics[width=3.2in]{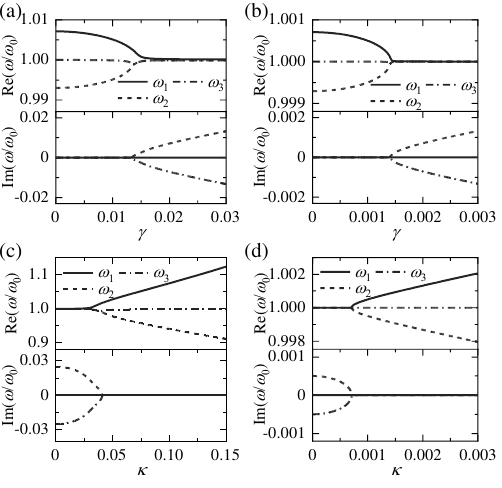}
    \caption{Eigenfrequency evolution under small gain-loss parameter and weak coupling condition when (a) $\kappa = 0.01$, (b) $\kappa = 0.001$, (c) $\gamma = 0.05$ and (d) $\gamma = 0.001$.}
    \label{fig:figure14}
\end{figure}

In both chained and planar-coupled structures within high-dimensional PT-symmetric circuit systems, the formation of higher-order exceptional points (HOEPs), where eigenfrequencies together with corresponding eigenmodes degenerate, may not necessarily occur. This stands in contrast to high-dimensional PT-symmetric optical systems where higher-order EPs can always be observed \cite{schnabel2017pt,heiss2016model}. In fact, in electronic systems, higher-order EPs tend to manifest predominantly when the coupling parameter $\kappa$ and the gain-loss parameter $\gamma$ are relatively small. It can be approximated that the system demonstrates a state of multiple modal degeneracies, and as the values of $\kappa$ and $\gamma$ approach zero, this approximation becomes more accurate, highlighting the dissimilarities between electronic circuits and optical systems. Taking a third-order PT-symmetric circuit with chained coupling as an example, \figref{fig:figure14} illustrates the evolution of the real and imaginary parts of the system's eigenfrequencies as a function of $\gamma$ when $\kappa = 0.01$ and $\kappa = 0.001$ respectively. 

As depicted in \figref{fig:figure14}(a), when $\kappa = 0.01$, the phase transition point still notably exhibits a twofold degeneracy (degeneracy of $\omega_2$ and $\omega_3$). When $\kappa$ decreases further to 0.001, it can be approximated that at this phase transition point, the modes corresponding to $\omega_{1,2,3}$  degenerate, suggesting an approximate third-order EP, as demonstrated in \figref{fig:figure14}(b). Additionally, with diminishing values of both the gain-loss parameter $\gamma$ and the coupling parameter $\kappa$, the real parts of the eigenfrequencies gradually converge towards one, approximating $\omega \approx \omega_0$. \figref{fig:figure14}(c) and \figref{fig:figure14}(d) compare the real and imaginary parts of the system's eigenfrequencies as a function of $\kappa$ for $\gamma = 0.05$ and $\gamma=0.001$, respectively. Similarly, when $\gamma = 0.05$, the phase transition points in the eigenfrequency evolution concerning $\kappa$ yield a twofold degeneracy (degeneracy of $\omega_2$ and $\omega_3$). However, as $\gamma$ further decreases to 0.001, it can be approximated that the phase transition point indicates a third-order EP where $\omega_{1,2,3}$ simultaneously degenerate.

Here, we explain the aforementioned phenomena from a mathematical modeling perspective. When both $\gamma$ and $\kappa$ are relatively small and the eigenfrequency $\omega$ is approximately equal to $\omega_0$, one can utilize the commonly used coupled-mode theory in optics to approximate the circuit theory. In such cases, the eigenfrequencies obtained from both theories align closely. In fact, the equations describing the system under coupled-mode theory are more concise, with lower-order equations that are easier to be discussed analytically. This facilitates the mathematical analysis of the conditions for the formation of higher-order EPs through expressions.

Considering three-dimensional chained parallel PT symmetric circuit illustrated in \figref{fig:figure02}, the system equation can be written in a matrix form $\bm{G} \bm{u} = 0$ where $\bm{u} = (u_1, u_2, u_3)^\text{T}$ and the coefficient matrix writes
\begin{equation} \label{eq:coeff_mat}
    \bm{G} = 
    \begin{pmatrix}
         -\cfrac{1}{R} + \iomega C + \cfrac{1-\kappa^2}{\iomega L\widetilde{\kappa}}  & -\cfrac{\kappa}{\iomega L\widetilde{\kappa}}         &  \cfrac{\kappa^2}{\iomega L\widetilde{\kappa}} \\[0.5em]
         -\cfrac{\kappa}{\iomega L\widetilde{\kappa}}                                &  \iomega C + \cfrac{1}{\iomega L\widetilde{\kappa}}  &  -\cfrac{\kappa}{\iomega L\widetilde{\kappa}} \\[0.5em]
         \cfrac{\kappa^2}{\iomega L\widetilde{\kappa}}                               & -\cfrac{\kappa}{\iomega L\widetilde{\kappa}}         &  \cfrac{1}{R} + \iomega C + \cfrac{1-\kappa^2}{\iomega L\widetilde{\kappa}} \\[0.5em]
    \end{pmatrix},
\end{equation}
where $\widetilde{\kappa} = 1 - 2 \kappa^2$. In the weak coupling regime where $\kappa^2 \approx 0$ and under approximation conditions $\omega \approx \omega_0$, the coefficient matrix \eqref{eq:coeff_mat} can be simplified to
\begin{equation} \label{eq:coeff_mat_simplified}
    \bm{G} \approx 
    \begin{pmatrix}
         -\cfrac{\gamma}{2}\omega_0 + \im(\omega-\omega_0)  & -\im \cfrac{\omega_0^2}{2\omega}\kappa & 0 \\
         -\im \cfrac{\omega_0^2}{2\omega}\kappa             & \im (\omega-\omega_0)                 &  -\im \cfrac{\omega_0^2}{2\omega}\kappa \\
         0                                                 & -\im \cfrac{\omega_0^2}{2\omega}\kappa & \cfrac{\gamma}{2}\omega_0 + \im(\omega-\omega_0)
    \end{pmatrix}.
\end{equation}
Detailed system equation derivation and approximation treatment are given in \appref{sec:sec:appendA}. Utilizing $\Exp^{\iomega t}$ time dependency and considering $u_n \propto \Exp^{\iomega t}$, \eqref{eq:coeff_mat_simplified} can be rewritten to Sch\"{o}dinger equation formalism as 
\begin{equation}
    \im \frac{\text{d}}{\text{d} t}
    \begin{pmatrix}
        u_1 \\
        u_2 \\
        u_3
    \end{pmatrix}
    =     
    \begin{pmatrix}
         -\omega_0 + \im \cfrac{\gamma}{2}\omega_0  & -\cfrac{\omega_0^2}{2\omega}\kappa & 0 \\
         -\cfrac{\omega_0^2}{2\omega}\kappa         & -\omega_0                         & -\cfrac{\omega_0^2}{2\omega}\kappa \\
         0                                         &  \cfrac{\omega_0^2}{2\omega}\kappa & -\omega_0 - \im \cfrac{\gamma}{2}\omega_0
    \end{pmatrix}
    \begin{pmatrix}
        u_1 \\
        u_2 \\
        u_3
    \end{pmatrix}.
\end{equation}
Since $\omega/\omega_0 \approx 1$ and $\widetilde{\gamma} = \omega_0 \gamma/2$ and $\widetilde{\kappa} = \kappa \omega_0/2$, we have $\im \text{d}\bm{u}/{\text{d} t} = \mathcal{H} \bm{u}$, where
\begin{equation} 
    \mathcal{H}=   
    \begin{pmatrix}
         -\omega_0 + \im \widetilde{\gamma}  & -\widetilde{\kappa}  & 0 \\
         -\widetilde{\kappa}                 & -\omega_0        & -\widetilde{\kappa} \\
         0                               & -\widetilde{\kappa}  & -\omega_0 - \im \widetilde{\gamma}
    \end{pmatrix}
\end{equation}
is the effective Hamiltonian. Solving the characteristic equation $\det(\mathcal{H}-\omega \identitymat_3) = 0$, the eigenvalues can be obtained as
\begin{subequations}
\begin{align}
    \omega_1 &=  \omega_0, \\
    \omega_{2,3} &=  \omega_0\left(1 \pm \sqrt{2\kappa^2-\gamma^2}\right).
\end{align}
\end{subequations}
It can be observed that when $\gamma < \sqrt{2}\kappa$, the eigenfrequencies consist of three distinct real numbers, indicating the system is in PT-symmetric phase. When $\gamma > \sqrt{2}\kappa$, the eigenfrequencies include one real number and a pair of complex conjugates, suggesting the system is in PT-broken phase. At the phase transition point where $\gamma = \sqrt{2}\kappa$, the eigenfrequencies exhibit a threefold degeneracy, all being $\omega_0$. This signifies that at this phase transition point, a third-order EP occurs. 

To quantitatively describe the accuracy of the coupled-mode equation, we define the mean deviation of three eigenvalues calculated by circuit theory as $D = \frac{1}{3} \sum_{j=1}^3 \text{Re}(\nomega_j)$. When using a coupled-mode equation to describe the system, the condition for the third-order EP will be $\kappa = \gamma/\sqrt{2}$. Substituting it into a circuit-theory-based equation to derive the accurate eigenfrequency, the result shows that: (1) For $D \leq 1\%$ the coupling coefficient should satisfy $\kappa \leq 0.05$; (2) For $D \leq 0.1\%$ the coupling coefficient should satisfy $\kappa \leq 0.009$. When using circuit theory to describe the system, the condition for EP will not be $\gamma = \sqrt{2} \kappa$ anymore. Here, we propose the definition of modified EP of three-dimensional PT-symmetric coupled $RLC$ resonators, which can be obtained by solving the discriminant $p^2 + q^3 = 0$ with $p$ and $q$ given in \eqref{eq:cardano_formula}. In this case, it can be evaluated that: (1) For $D \leq 1\%$, the gain-loss parameter and the coupling coefficient should satisfy $\kappa \leq 0.037$; (2) For $D \leq 0.1\%$, the gain-loss parameter and coupling coefficient should satisfy $\kappa \leq 0.0064$.

\section{Conclusion}
This article theoretically investigates the non-Hermitian properties of typical three-dimensional linear and planar-coupled PT-symmetric electronic resonators. Starting from system equations based on circuit theory, the non-Hermitian phase transition characteristics from PT-symmetric to symmetry-broken phase are verified respectively. The result indicates that for parallel topology, the PT-symmetric phase typically corresponds to strong coupling regime and small gain-loss parameters $\gamma$, with voltage waveforms exhibiting Rabi oscillations; while in the PT-broken phase the system exhibits exponentially divergent voltage waveforms, showing under-damped and over-damped response pattern, respectively, as $\gamma$ increases. In an $n$-dimensional system, the voltage waveform contains $n$ eigenfrequency components. Furthermore, by comparing circuit theory with coupled-mode theory, the conditions for the formation of high-order EPs in high-dimensional electronic circuits are analyzed. When both the coupling and gain-loss parameters are significantly smaller than one, and the eigenfrequencies are close to the natural resonant frequencies of the $LC$ resonators constituting the system, the description based on coupled-mode theory can approximate the rigorous description of circuit theory well. At this point, non-Hermitian circuits can be analogous to non-Hermitian optical systems based on coupled-mode descriptions, sharing similar properties. In the strong coupling regime, electronic circuits may exhibit richer characteristics such as divergent singular points, whose potential application values are yet to be fully explored and exploited.

\begin{acknowledgments}
K.Y. acknowledges the support of the National Natural Science Foundation of China (52407015) and the Postdoctoral Fellowship Program of CPSF (GZB20240469). K.T. acknowledges the support of the Natural Science Foundation of Sichuan Province (2023NSFSC1426). T.D. acknowledges the support of Key Research and Development Program of Shaanxi Province (2024GX-YBXM-236).
\end{acknowledgments}

\section*{Author Declarations}
The authors have no conflicts to disclose.

\section*{Data Availability Statement}
The data that support the findings of this study are available from the corresponding author upon reasonable request.

\appendix
\section{Derivation of System Equation}
For the circuit schematic shown in \figref{fig:figure07}, according to Kirchoff's current law (KCL), we have
\begin{subequations} \label{eq:kcl}
    \begin{align}
        i_1 - \frac{u_1}{R} + C \frac{\text{d} u_1}{\text{d} t} &= 0, \\
        i_2 + C \frac{\text{d} u_2}{\text{d} t} &= 0, \\
        i_3 + \frac{u_3}{R} + C \frac{\text{d} u_3}{\text{d} t} &= 0,
    \end{align}
\end{subequations}
where $u_n$ and $i_n$ represent capacitor's voltage and inductor's current respectively; $R$, $L$, and $C$ represent the values of the resistance, inductance, and capacitance of the $RLC$ resonator, respectively. Furthermore, based on the $i-v$ relationship of an inductor, we can obtain
\begin{subequations} \label{eq:inductor-vcr}
    \begin{align}
        u_1 &= L \frac{\text{d} i_1}{\text{d} t} + M_1 \frac{\text{d} i_2}{\text{d} t} + M_2 \frac{\text{d} i_3}{\text{d} t}, \\
        u_2 &= L \frac{\text{d} i_2}{\text{d} t} + M_1 \left(\frac{\text{d} i_1}{\text{d} t} + \frac{\text{d} i_3}{\text{d} t}\right), \\
        u_3 &= L \frac{\text{d} i_3}{\text{d} t} + M_1 \frac{\text{d} i_2}{\text{d} t} + M_2 \frac{\text{d} i_1}{\text{d} t},
    \end{align}
\end{subequations}
where $M_1$ and $M_2$ represent the mutual inductance between gain-neutral inductors and neutral-loss inductors respectively. Taking $u_1$, $u_2$ and $u_3$ as system variables and eliminating $i_1$, $i_2$ and $i_3$ in \eqref{eq:inductor-vcr}, we can obtain the system equation given in \eqref{eq:system_equation_pt3_planar}. The linear configuration system equation given in \eqref{eq:system_equation_linear} is just the special case of \eqref{eq:system_equation_pt3_planar} when $M_1 = M$ and $M_2 = 0$.

\section{Derivation of Coupled-Mode Equation} \label{sec:sec:appendA}
For three-dimensional chained parallel PT symmetric circuit illustrated in \figref{fig:figure02}, the relationship between voltage and current in the inductors of each resonator can be written as
\begin{equation}
    \iomega
    \begin{pmatrix}
         L & M & 0 \\
         M & L & M \\
         0 & M & L
    \end{pmatrix}
    \begin{pmatrix}
        i_1 \\
        i_2 \\
        i_3
    \end{pmatrix}
    =
    \begin{pmatrix}
        u_1 \\
        u_2 \\
        u_3
    \end{pmatrix}.
\end{equation}
Accordingly, $(i_1, i_2, i_3)^\text{T}$ can be derived as
\begin{equation} \label{eq:induct_mat}
    \begin{pmatrix}
        i_1 \\
        i_2 \\
        i_3
    \end{pmatrix}
    = \frac{1}{\iomega L(L^2-2M^2)}
    \begin{pmatrix}
         L^2-M^2 & -LM & M^2 \\
         -LM     & L^2 & -LM \\
         M^2     & -LM &  L^2-M^2
    \end{pmatrix}
    \begin{pmatrix}
        u_1 \\
        u_2 \\
        u_3
    \end{pmatrix}.
\end{equation}
According to Kirchhoff's law, $i_1$, $i_2$ and $i_3$ can also be written in matrix form as 
\begin{equation} \label{eq:i1i2i3}
    \begin{pmatrix}
        i_1 \\
        i_2 \\
        i_3
    \end{pmatrix}
    =
    \begin{pmatrix}
         \cfrac{1}{R}-\iomega C &     0       & 0 \\
         0                     &  -\iomega C & 0 \\
         0                     &     0       &  \cfrac{1}{R} - \iomega C
    \end{pmatrix}
    \begin{pmatrix}
        u_1 \\
        u_2 \\
        u_3
    \end{pmatrix}.
\end{equation}
Substitute \eqref{eq:i1i2i3} into \eqref{eq:induct_mat} and let $M = \kappa L$, we can obtain the system equation in matrix form $\bm{G} \bm{u} = 0$ with $\bm{G}$ given in \eqref{eq:coeff_mat}. In the weak coupling regime, $\kappa ^2 \approx 0$, the system equation can be approximated as
\begin{equation}
    \begin{pmatrix}
         -\cfrac{1}{R} + \iomega C + \cfrac{1}{\iomega L}  & -\cfrac{\kappa}{\iomega L}  & 0  \\
         -\cfrac{\kappa}{\iomega L} &  \iomega C + \cfrac{1}{\iomega L}  & -\cfrac{\kappa}{\iomega L}  \\
         0 & -\cfrac{\kappa}{\iomega L} & \cfrac{1}{R} + \iomega C + \cfrac{1}{\iomega L} 
    \end{pmatrix}
    \begin{pmatrix}
        u_1 \\
        u_2 \\               
        u_3
    \end{pmatrix}
    = 0.
\end{equation}
Defining $\gamma = R^{-1}\sqrt{L/C}$, we can obtain
\begin{equation} \label{eq:u123omega}
    \begin{pmatrix}
         -\gamma + \im \overline{\omega}  & -\im \cfrac{\kappa \omega_0}{\omega}  &   0   \\
         -\im \cfrac{\kappa \omega_0}{\omega} & \im \overline{\omega}  & -\im \cfrac{\kappa \omega_0}{\omega} \\
         0 & -\im \cfrac{\kappa \omega_0}{\omega} & \gamma + \im \overline{\omega}
    \end{pmatrix}
    \begin{pmatrix}
        u_1 \\
        u_2 \\
        u_3
    \end{pmatrix}
    = 0,
\end{equation}
where $\overline{\omega} = \omega/\omega_0 - \omega_0/\omega$ and $\omega_0 = 1/\sqrt{LC}$ is the natural frequency of the $LC$ resonator. Considering $\overline{\omega} = \omega/\omega_0 - \omega_0/\omega = (\omega^2 - \omega_0^2)/(\omega \omega_0) = [(\omega+\omega_0)(\omega-\omega_0)]/(\omega \omega_0) = (1 + \omega_0/\omega) (\omega - \omega_0)/\omega_0 \approx 2 (\omega - \omega_0)/\omega_0$, multiply $\omega_0/2$ to \eqref{eq:u123omega}, we can obtain system equation with simplified coefficient matrix given in \eqref{eq:coeff_mat_simplified}. Finally, the system equation can be rewritten to Sch\"{o}dinger equation formalism as 
\begin{equation}
    \im \frac{\text{d}}{\text{d} t}
    \begin{pmatrix}
        u_1 \\
        u_2 \\
        u_3
    \end{pmatrix}
    =     
    \begin{pmatrix}
         -\omega_0 + \im \widetilde{\gamma}  & -\widetilde{\kappa}  & 0 \\
         -\widetilde{\kappa}                 & -\omega_0        & -\widetilde{\kappa} \\
         0                               & -\widetilde{\kappa}  & -\omega_0 - \im \widetilde{\gamma}
    \end{pmatrix}
    \begin{pmatrix}
        u_1 \\
        u_2 \\
        u_3
    \end{pmatrix}.
\end{equation}

\bibliography{main}

\end{document}